%% file: JetML.tex
\documentclass[a4paper,11pt]{article}
\pdfoutput=1 

\usepackage{jheppub} 

\usepackage[T1]{fontenc} 
\usepackage{siunitx}
\usepackage{lineno}
\usepackage[perpage]{footmisc}                     
\usepackage[utf8]{inputenc}
\usepackage{graphicx} 
\usepackage{grffile}
\usepackage{subcaption}
\usepackage{dcolumn}  
\usepackage{makecell}
\usepackage{colordvi}
\usepackage{color}
\usepackage{epstopdf}
\usepackage{amssymb}
\usepackage{amsmath}
\usepackage{bm}
\usepackage{slashed}
\usepackage{enumerate}
\usepackage{url}
\usepackage{morefloats}
\usepackage{dirtytalk}
\usepackage{natbib}
\usepackage[T1]{fontenc} 
\usepackage{siunitx}
\usepackage{graphicx}
\usepackage{lineno}
\usepackage[perpage]{footmisc}
\usepackage{comment}
\usepackage[colorlinks=true,
urlcolor=blue,
linkcolor=blue,
citecolor=blue,
linktocpage=true,
pdfproducer=medialab,
pdfa=true,
anchorcolor=blue]{hyperref}

\usepackage{tabularx,array}

\newcolumntype{C}{>{\centering\arraybackslash}X}
\usepackage{multirow}

\usepackage{lineno}

\usepackage{float}
\usepackage{xcolor}

\usepackage[title]{appendix}

\renewcommand{\thesection}{\arabic{section}}

\usepackage{changepage}
\usepackage{booktabs}


\preprint{}

\begin{document}

\title{\boldmath Validating a Machine Learning Approach to Identify Quenched Jets in Heavy-Ion Collisions
}

\author[a]{Yilun Wu,}
\author[a]{Yi Chen,}
\author[a]{Julia Velkovska}

\affiliation[a]{Vanderbilt University, Nashville, Tennessee, USA}

\emailAdd{yilun.wu@vanderbilt.edu}
\emailAdd{luna.chen@vanderbilt.edu}
\emailAdd{julia.velkovska@vanderbilt.edu}

\date{\today}

\abstract{
Jet quenching is a phenomenon in heavy-ion collisions arising from jet interactions with the quark-gluon plasma (QGP). Its study is complicated by the interplay of multiple physics processes that affect jet observables. In addition, detector effects may influence the results and must be accounted for when identifying quenched jets. We employ a Long Short-Term Memory (LSTM) neural network trained on jet substructure, incorporating parton shower history, to predict jet-by-jet quenching levels. Using photon-jet samples from the \textsc{Jewel} event generator, we show that the LSTM predictions strongly correlate with true jet energy loss. This validates that the model effectively learns the features of jet-QGP interaction. We simulate detector effects using \textsc{Delphes} simulation framework and demonstrate that the method identifies quenching effects in a realistic environment. We test the approach with photon-jet momentum imbalance, jet fragmentation function, and jet shape, which were not included in the training, confirming its ability to distinguish true quenching features. 

}

\maketitle
\flushbottom


\section{Introduction}
\label{sec:Introduction}

Jets, collimated sprays of high‐\(p_\mathrm{T}\) hadrons produced from the fragmentation of hard-scattered partons, serve as probes of the quark–gluon plasma (QGP) \cite{Bjorken:1982tu}. The phenomenon of jet quenching, first evidenced by the suppression of high‐\(p_\mathrm{T}\) hadrons at the Relativistic Heavy Ion Collider (RHIC) \cite{Adcox:2001jp,STAR:2002ggv,Adler:2003qi,Adams:2003kv,STAR:2002svs} and later confirmed at the Large Hadron Collider (LHC) \cite{Aamodt:2010jd,Aamodt:2011vg,CMS:2012aa,Aad:2015wga}, indicates the formation of QGP.

In heavy ion collisions, interactions with the QGP medium modify the total energy and the internal structure of the jets as compared to proton-proton collisions. Jet suppression ~\cite{CMSRAA, Aad:2012vca,Abelev:2013kqa, Adam:2015ewa} depends on factors such as jet flavor~\cite{CMS:2013qak, Aad_jetflavr2023, Aad_jetmass2023}, path-length~\cite{Atlasjetpathl, CMSjetpathl} and fluctuations in medium density \cite{FluctQGP, FluctQGP2023}. To understand how lost energy is redistributed, studies have focused on jet-hadron correlations, such as jet shapes~\cite{Chatrchyan:2013kwa,ALICE:2019whv,ATLAS:2019pid}, and fragmentation functions~\cite{Chatrchyan:2014ava,Aaboud:2017bzv}. Beyond these, jet substructure measurements, including groomed jet observables~\cite{Sirunyan:2017bsd, STAR:2021kjt,ALargeIonColliderExperiment:2021mqf, CMSjetmg, aad2023measurement} and energy-energy correlators~\cite{CMS:2024ovv}, offer insights into jet evolution from parton showering to hadronization. Jet substructures are sensitive to the QGP resolution scale  and color coherence~\cite{aad2023measurement}, providing new perspectives on parton energy loss.

Disentangling these factors remains challenging. The experimental selections tend to have bias towards jets that lose less energy~\cite{CMSrg_2025}. In addition, energy fluctuations from residual uncorrelated background particles~\cite{Connors_2018, Mulligan:2020tim}, detector resolution, and differences in the relative fraction of quark and gluon jets between heavy-ion and pp collisions~\cite{Ringer_2020, 2025flavordependencEEC} may also appear as modifications. Knowing the jet energy loss on a jet-by-jet basis could greatly improve our understanding of jet-QGP interactions. In recent years, machine learning has proven to be effective in identifying individual jets based on their degrees of modification by the QGP \cite{Apolinario:2021olp, Du:2020pmp, Lai:2021ckt, Crispim_Rom_o_2024}, even when accounting for simplified thermal background \cite{liu2023identifying, Goncalves:2025fpf} and detector smearing \cite{qureshi2024modelagnostic}. While most simulation-based studies achieve strong discrimination between jets from pp and PbPb collisions, it is a non-trivial task to cross-check whether classifiers distinguish true quenching features or over-learn from other effects that mimic quenching. 

In this study, we train a Long Short-Term Memory (LSTM) network~\cite{hochreiter1997long,Sherstinsky_2020} using supervised learning to distinguish quenched jets from vacuum jets on an individual basis. Our approach incorporates a realistic uncorrelated background along with the full detector response of both the calorimeter and the tracker. The resulting machine learning classifier employs the jet substructure and features of the parton shower history to preserve essential information about the parton energy loss in the QGP. We confirm that the quenching predictions from the LSTM strongly correlate with jet energy loss tagged by photons, even in the presence of underlying events and realistic detector responses. The trained classifier is then applied to observables not included in the training process as a cross-check to validate its ability to distinguish true quenching features from unrelated effects that mimic quenching. 

The paper is organized as follows. In Section~\ref{sec:Samples}, we describe the event simulation procedure, as well as the detector response and correction methods. Section~\ref{sec:Jet reconstruction} discusses the jet reconstruction and energy correction process at the detector level. 
Section~\ref{sec:Machine learning setup} details the supervised machine learning approach, including the input to it being optimized to capture jet energy loss information. In Section~\ref{sec:Results}, we validate the quenching predictions of the trained classifier using various observables and identify medium jets with different levels of modification. In Section~\ref{sec:Conclusion}, we present our conclusions and outlook. 

\section{Simulated samples}
\label{sec:Samples}

\subsection{Event generation}
Photon-jet samples are generated using \textsc{Jewel v2.2.0}~\cite{Zapp:2012ak,KunnawalkamElayavalli:2017hxo} with the "PPYJ" process at $\sqrt{s_{\mathrm{NN}}} = \SI{5.02}{TeV}$ and a minimum transverse momentum of $\hat p_{\mathrm{T}} = \SI{40}{GeV}$. For heavy-ion jets, we enable the "simple" option with "recoil" on to include parton-medium interactions and medium response, while pp jets are simulated with the "vacuum" option. All \textsc{Jewel} events used in this analysis are generated in unweighted mode, and no event weights are applied in neural network training.

\subsection{Treatment of JEWEL recoil subtraction}
For \textsc{Jewel} events generated with recoil enabled, a \textsc{Jewel}-specific subtraction is required to remove the thermal momentum component carried by recoil partons. In this work,  we implement the Thermal Momenta Subtraction procedure~\cite{Milhano:2022kzx}, applied at the generator level prior to embedding into a thermal background \cite{Goncalves:2025fpf}. Recoil contributions are removed using a matching criterion of $\Delta R_{ik} < 0.5$, ensuring that unphysical recoil components do not contribute to the features used for machine learning.The classifier performance is found to be insensitive to this procedure.

\subsection{Thermal background and subtraction}
Both medium and vacuum jet samples are embedded in a realistic heavy-ion background. After embedding, the same constituent subtraction procedure~\cite{Berta:2014eza} is applied to both medium and vacuum jet samples to remove the soft background and prevent the neural network from learning uncorrelated features that could mimic quenching effects~\cite{Goncalves:2025fpf}. 

The background simulation uses \textsc{Angantyr}~\cite{Bierlich_2018} within \textsc{Pythia 8.3}~\cite{SciPostPhysCodeb.8} for 0--10\% central heavy-ion events at $\sqrt{s_{\mathrm{NN}}} = \SI{5.02}{TeV}$. The outgoing partons have a maximum transverse momentum of $\hat{p}_{\mathrm{T}} = \SI{5}{GeV}$\footnote{For more information, see main42.cc and main42.cmnd in \textsc{Pythia 8.3} examples.}.  Although \textsc{Angantyr} does not include a hydrodynamic description of QGP evolution, it provides a realistic description of the bulk properties of heavy-ion events, such as multiplicity and momentum density, which dominate the background fluctuations relevant for jet reconstruction. In this work, the underlying event is treated as an uncorrelated background, while medium-induced modifications are modeled explicitly by \textsc{Jewel}. 

In Figure~\ref{fig:BkgMultiplicity}, the average momentum density for each simulated underlying event is correlated with the multiplicity and is found to be consistent with CMS measurements~\cite{Hayrapetyan:2897965}. The average momentum density is also consistent with our previous study~\cite{VandyMLJets}, with the present approach additionally capturing event-by-event fluctuations. More details are listed in Table~\ref{tab:tab1}. 

We therefore do not rely on \textsc{Angantyr} to describe medium response, but only to model the underlying event fluctuations relevant for jet reconstruction.

\begin{figure}[h]
        \centering
        
        \includegraphics[width=0.5\textwidth]{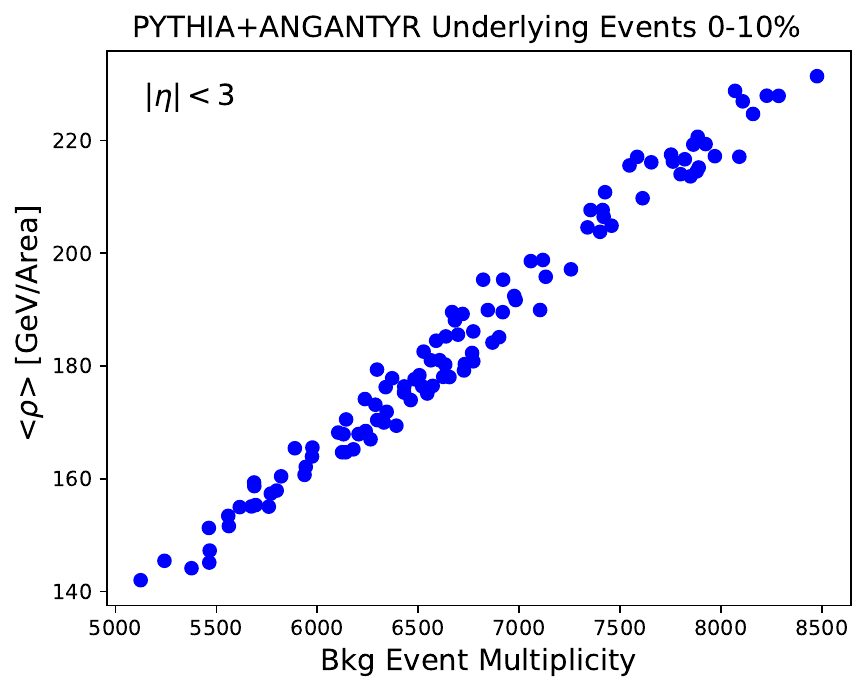}
        \label{fig:BkgMultiplicity}       
        \caption{Scatter plot of average momentum density $\langle \rho \rangle$ and multiplicity for the underlying event simulated by \textsc{Angantyr} within \textsc{Pythia 8.3}. }
        \label{fig:BkgMultiplicity}
\end{figure}

\begin{table}[ht]
\begin{center}
\begin{tabular}{ c|c } 
 Values & \textsc{Angantyr} \\ 
 \hline
 Pseudorapidity Interval  & $|\eta|<3$ \\ 
 Mean Event Multiplicity  & 6727 \\
 Average Yield $\langle \frac{\mathrm{d}N}{\mathrm{d}\eta} \rangle$  & 1121  \\
 Average Transverse Momentum $\langle p_{\mathrm{T}} \rangle$  & $\SI{1.05} {GeV}$\\
 Average Transverse Momentum Density $\langle \rho \rangle$ & $\SI {185.0} {GeV/Area}$\\
 \hline
\end{tabular}
\bigskip
\caption{Parameters of the simulated underlying events. }
\label{tab:tab1}
\end{center}
\end{table}
\subsection{Event selection}
The machine learning approach is applied to photon-jet events, where the photon energy provides an independent calibration of the jet energy loss. Following experimental practice, we select \textsc{Jewel} events in which the highest-$p_T$ photon and highest-$ p_T $ jet in the event have an angular separation of $ 7\pi/8 $. The jets selected for training (learning set) have back-to-back photons with $ p_T \in [\SI{200}, \SI{300}] \text{~GeV}$, jet $ p_T > \SI{40}{GeV}$, and $ |\eta| < 2.0 $. This photon energy cut during training constrains the neural network, ensuring that it learns primarily from energy loss rather than fragmentation bias. However, a well-trained model applied to jet quenching classification across different observables (classification set) can adapt to a photon energy range without an upper limit (e.g., $ p_T^\gamma > \SI{200}{GeV}$). 

The reconstructed jets are reclustered using the Cambridge/Aachen (C/A) algorithm ~\cite{Dokshitzer:1997in} to extract jet substructure observables. The C/A algorithm organizes the jet constituents into an angularly ordered clustering tree. Within this structure, significant splittings are identified using the iterative soft-drop grooming procedure~\cite{Larkoski:2014wba}, which removes soft, wide-angle branches while retaining those that satisfy the soft-drop condition, 

\begin{equation}
    z_{\mathrm{g}} \equiv \frac{\mathrm{min}(p_{\mathrm{T}1} , p_{\mathrm{T}2})}{p_{\mathrm{T}1}+p_{\mathrm{T}2}} > z_{\mathrm{cut}} \left( \frac{\Delta R}{R_0} \right) ^\beta,
    \label{eq:sd}
\end{equation}
where $p_{\mathrm{T1}}$ and $p_{\mathrm{T2}}$ are the transverse momenta of the two distinct subjet branches; $\Delta R$ represents the angular separation between them and $R_0$, set to 0.4, is the jet radius in this study. We set the parameters to $z_{\mathrm{cut}} = 0.1$ and $\beta = 0$. Table~\ref{tab:tab2} summarizes the cuts used in event selection. 

\begin{table}[h]
\begin{center}
\begin{tabular}{ c|c|c } 
 Photon-Jet Selection & Learning set & Classification set \\ 
 \hline
 Centrality Interval & 0-10\% & 0-10\% \\
 Jet Pseudorapidity Interval &$|\eta|<2.0$ & $|\eta|<2.0$\\
 Photon-Jet Angle Separation  & $7\pi/8$ & $7\pi/8$ \\
 Photon Energy  & [200,300]GeV  & >200 GeV\\ 
 Jet Energy & >40 GeV & >100 GeV \\
 Soft Drop Condition& $z_{\mathrm{cut}} = 0.1$, $\beta = 0$ & $z_{\mathrm{cut}} = 0.1$, $\beta = 0$ \\
 No. Vac-Jets & 38312 & 64074 \\
 No. Med-Jets & 39629 & 74580 \\
 
 \hline
\end{tabular}
\bigskip
\caption{Photon-jet selection cuts in training and classification.}
\label{tab:tab2}
\end{center}
\end{table}

\subsection{Detector fast simulation}
To study detector effects on the reconstruction of substructure observables and neural network performance, we use \textsc{Delphes-3.5.0} \cite{DELPHES3} to simulate CMS detector conditions. The tracking efficiency of charged particles depends on $\eta$ and $\phi$, with a momentum threshold of $\SI{0.55}{GeV}$. The particle energies are smeared according to the ECAL and HCAL calorimeter resolutions, matching those of the CMS detector. Additionally, we employ \textsc{Delphes} particle-flow emulation, following the CMS approach \cite{CMS-PAS-PFT-09-001}, optimally combining information from all subdetectors to reconstruct and identify individual particles.

\section{Jet reconstruction}
\label{sec:Jet reconstruction}

Jets are reconstructed using the anti-$k_\mathrm{T}$ algorithm~\cite{Cacciari:2008gp} from the emulated particle-flow candidates with a distance parameter of $R=0.4$. The underlying event particles are subtracted using the event-wide constituent subtraction method~\cite{Berta:2014eza}. The maximum distance parameter between the signal and the background particles is set to $\Delta R_{\mathrm{max}} = 0.3$, with a $p_{\mathrm{T}}$ weight of $\alpha=1$. 

\subsection{Jet energy scale and resolution}

\begin{figure}[h]
        \centering
        \begin{subfigure}[b]{0.48\textwidth}
            \centering
            \includegraphics[width=\linewidth]{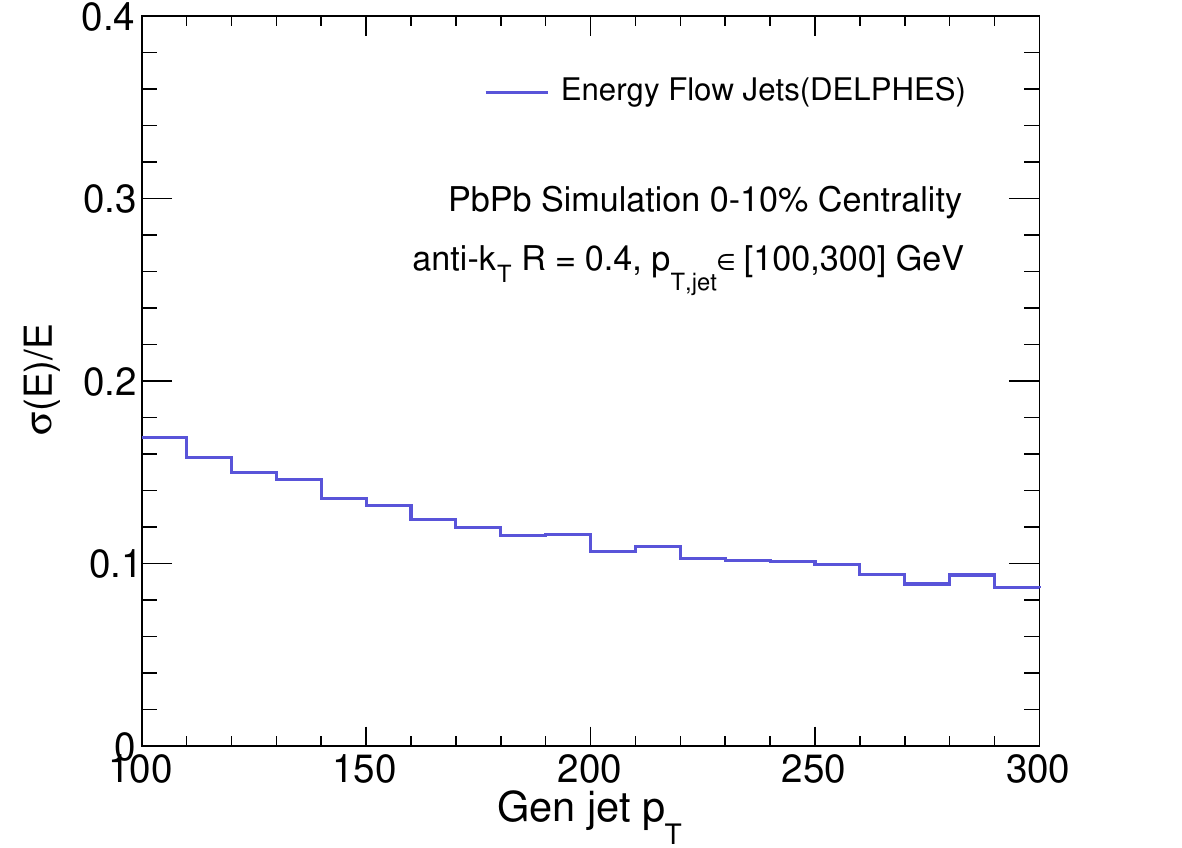}
            \caption{}
            \label{fig:JER_PbPb}
        \end{subfigure}
        \begin{subfigure}[b]{0.48\textwidth}
            \centering
            \includegraphics[width=\linewidth]{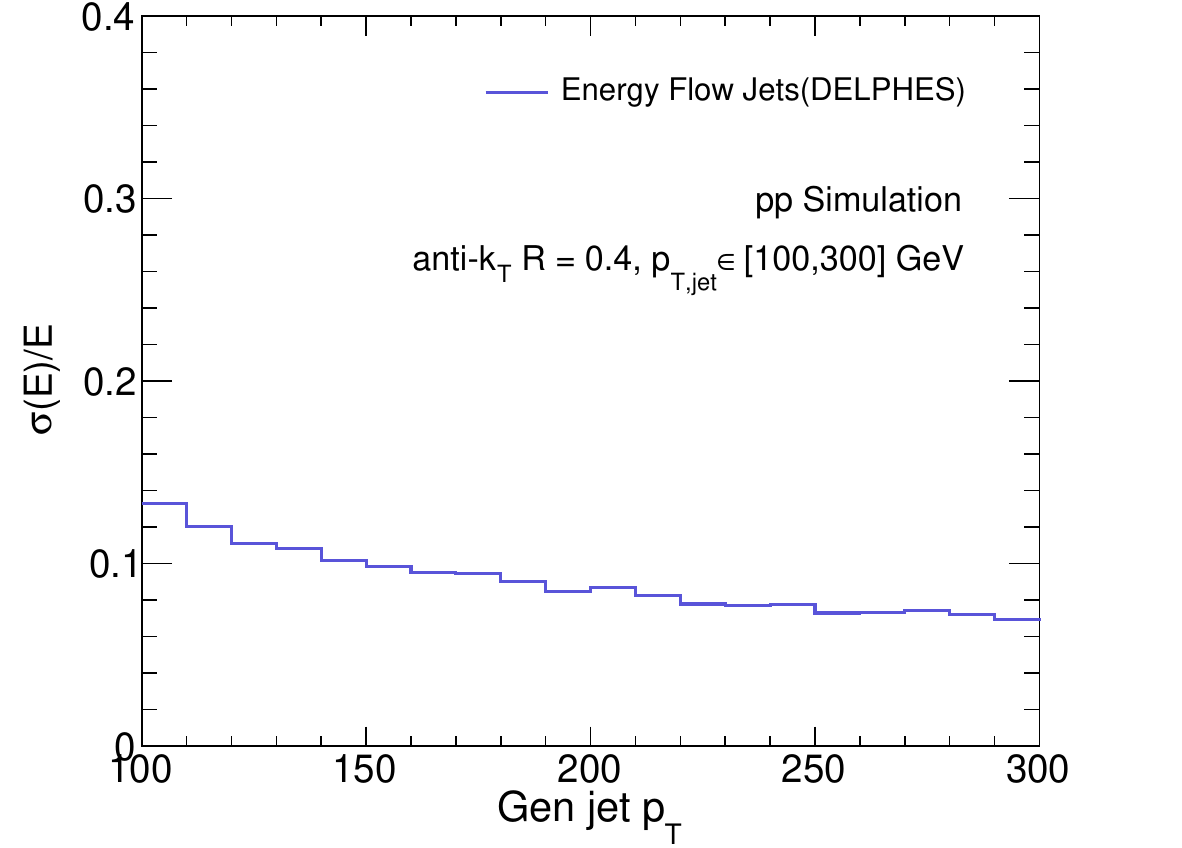}
            \caption{}
            \label{fig:JER_pp}
        \end{subfigure}
        \caption{JER of reconstructed medium/vacuum jets from \textsc{Delphes} Energy Flow candidates. (a): JER for medium jets; (b): JER for vacuum jets. }
        \label{fig:JER}
\end{figure}

The reconstructed jets are matched to their GEN-level counterparts to determine the relative energy scales. Figures \ref{fig:JESPbPb} and \ref{fig:JESpp} illustrate the jet energy scale (JES) for reconstructed medium and vacuum jets in simulated PbPb and pp collisions, respectively, with detector responses simulated by \textsc{Delphes}, compared to jets at GEN-level. The calibration of the jet energy recovers the jet energy from the RECO level to the GEN level as a function of $p_T$, $\eta$, and $\phi$. Figure \ref{fig:JER} shows the jet energy resolution (JER) for \textsc{Jewel}-Med (PbPb) and \textsc{Jewel}-Vac (pp) jets.

\begin{figure}[h]
        \centering

        \includegraphics[width=0.98\textwidth]{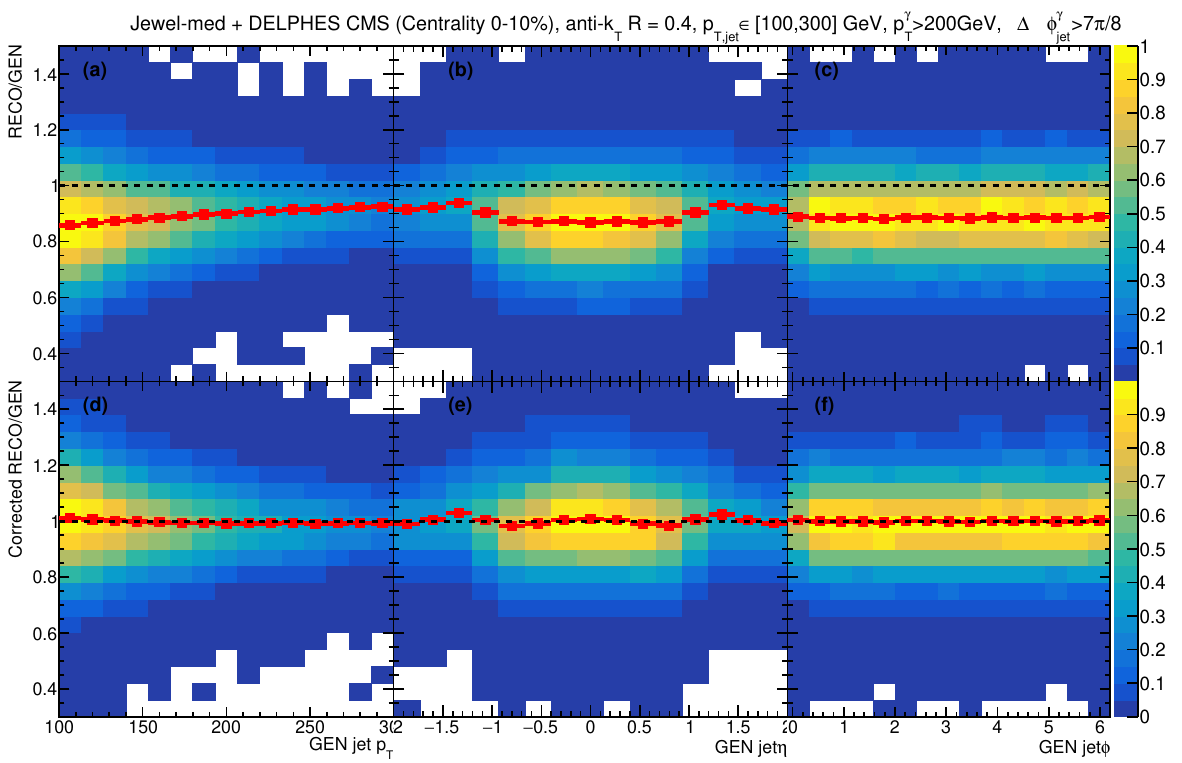}
        \label{fig:JESPbPb}

       \caption{JES of reconstructed \textsc{Jewel}-Med (PbPb) jets from \textsc{Delphes} Energy Flow candidates. The upper panel shows JES before jet energy calibration, while the lower panel shows JES after calibration. (a–c) JES as a function of jet \( p_{\mathrm{T}} \), \( \eta \), and \( \phi \), respectively. (d–f) Corresponding distributions after jet energy corrections.}

        \label{fig:JESPbPb}
\end{figure}

\begin{figure}[h]
        \centering
        \includegraphics[width=0.98\textwidth]{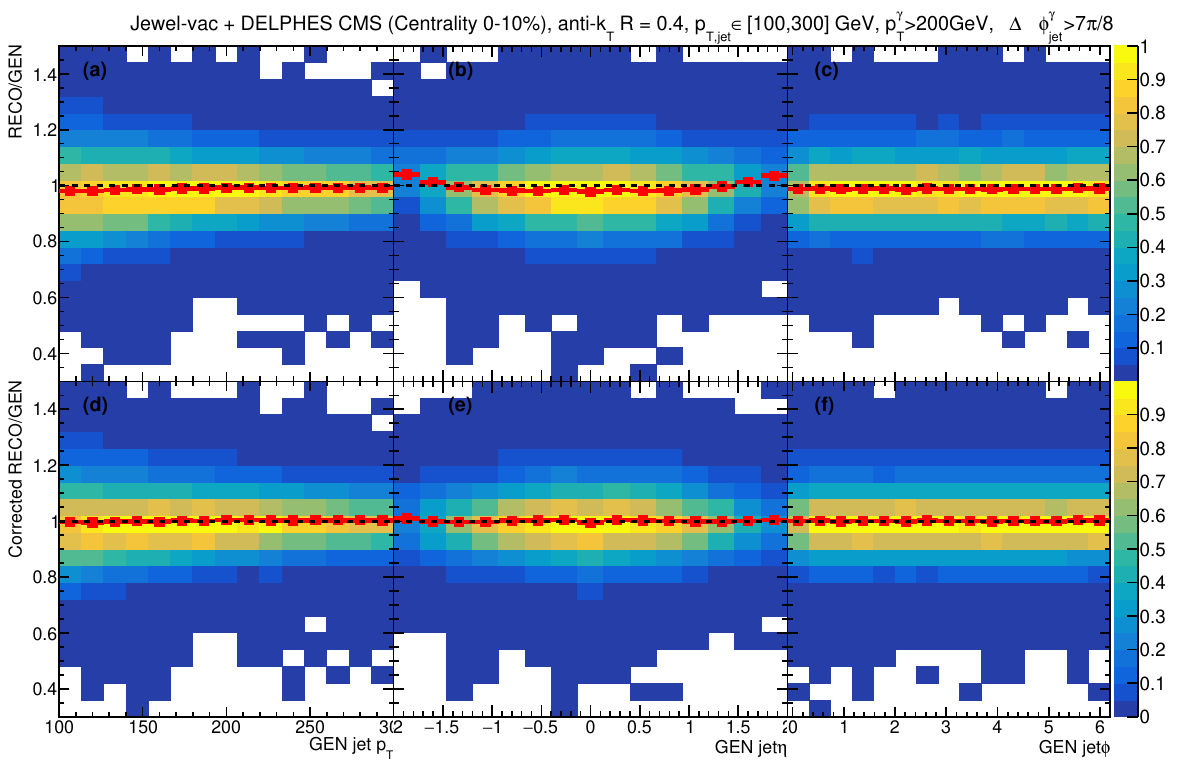}
        \label{fig:JESpp}       
        \caption{JES of reconstructed \textsc{Jewel}-Vac (pp) jets from \textsc{Delphes} Energy Flow candidates. The upper panel shows JES before jet energy calibration, while the lower panel shows JES after calibration. (a–c) JES as a function of jet \( p_{\mathrm{T}} \), \( \eta \), and \( \phi \), respectively. (d–f) Corresponding distributions after jet energy corrections.}

        \label{fig:JESpp}
\end{figure}

\clearpage

\section{Machine learning setup}
\label{sec:Machine learning setup}

Jets are declustered following angular-ordered structures (Sec.~\ref{sec:Jet reconstruction}), retaining the harder subjet as sequential input for machine learning. Each selected splitting is represented by a feature vector  $x_{\mathrm{t}}$, comprising the momentum fraction $z$, the angular distance $\Delta R$, the perpendicular momentum $k_\perp$ and the invariant mass $m_{\mathrm{inv}}$, derived from the declustered subjet pair:

\begin{equation}
\begin{aligned}
\centering
     & z = \frac{\mathrm{min}(p_{\mathrm{T,i}} , p_{\mathrm{T,j}})}{p_{\mathrm{T,i}}+p_{\mathrm{T,j}}},\\
     & \Delta R = \sqrt{(\phi_{\mathrm{i}}-\phi_{\mathrm{j}})^2+(\eta_{\mathrm{i}}-\eta_{\mathrm{j}})^2},\\
     & k_{\perp} = \mathrm{min}(p_{\mathrm{T,i}} , p_{\mathrm{T,j}}) * \Delta R,\\
     & m_{\mathrm{inv}} = \sqrt{(E_{\mathrm{i}}+E_{\mathrm{j}})^2-(\textbf{p}_{\mathrm{i}}+\textbf{p}_{\mathrm{j}})^2},\\
     & x_{\mathrm{t}} = [z, \Delta R, k_{\perp}, m_{\mathrm{inv}}],\\
\end{aligned}
\end{equation}
where $i, j$ denote the subjets in the declustering step $t$. In this way, sequential vectors $[x_0,...,x_{\mathrm{t}},...]$ are extracted and used in the training of a neural network. Sequential vectors, which encode the showering history of the jet while preserving modifications from QGP medium, serve as input to train a neural network composed of LSTM and fully connected (FC) layers implemented in PyTorch~\cite{NEURIPS2019_9015}. As a type of recurrent neural network, an LSTM is particularly suited to learn from sequential data, using internal "gates"~\cite{Sherstinsky_2020} to selectively retain or discard information over time steps. This allows it to effectively capture the patterns of jet-QGP interaction.

Jet quenching identification is framed as a classification problem of supervised learning, with vacuum jets labeled 0 and medium jets labeled 1. In the final step, predicted values for all input jets are used to compute batch losses via the weighted mean squared error (MSE) loss function:
\begin{equation}
\begin{aligned}
\centering
& l_{\mathrm{MSE}} = \frac{\sum\limits_{\mathrm{batch}} \omega_{\mathrm{i}} * (x_{\mathrm{i}} - y_{\mathrm{i}})^2}{\sum\limits_{\mathrm{batch}} \omega_{\mathrm{i}}},
\end{aligned}
\end{equation}
where  $x_{\mathrm{i}}$ and $y_{\mathrm{i}}$ denote the predicted and true labels of the $i$th jet, respectively, while $\omega_{\mathrm{i}}$ Model hyper-parameters, including the number of LSTM layers, fully connected (FC) layer dimensions\footnote{The input dimension of the first FC layer equals the LSTM hidden size.}, epochs, batch size, learning rate, and decay factor, are optimized using the Hyperopt package~\cite{pmlr-v28-bergstra13}. Following the same training method as in~\cite{liu2023identifying}, we perform 50 search iterations, training each configuration three times, and select the model with the lowest validation loss as the best-performing. The corresponding optimal hyper-parameters are listed in Table~\ref{tab:tab3}.

\begin{table}[ht]
\begin{center}
\begin{tabular}{ c|c|c } 
 Parameters & GEN Level & RECO Level\\ 
 \hline
 No. of LSTM Layers & 2 & 2\\ 
 FC Dim 1 & 14 & 14\\
 FC Dim 2 & 8 & 4 \\
 \hline
 No. of Epochs & 50 & 50\\
 Batch Size& 12000 & 10000\\
 Learning Rate & 0.04754 & 0.0416\\
 Decay Factor & 0.9728 & 0.9810\\
 \hline
\end{tabular}
\bigskip
\caption{List of hyper-parameters related to the neural network architecture and the training process, and their optimal values after hyper-tuning.}
\label{tab:tab3}
\end{center}
\end{table}

\section{Results}
\label{sec:Results}
\subsection{Training performance}
\label{Training performance}

\begin{figure}[h]
        \centering
        \begin{subfigure}[b]{0.48\textwidth}
            \centering
            \includegraphics[width=\linewidth]{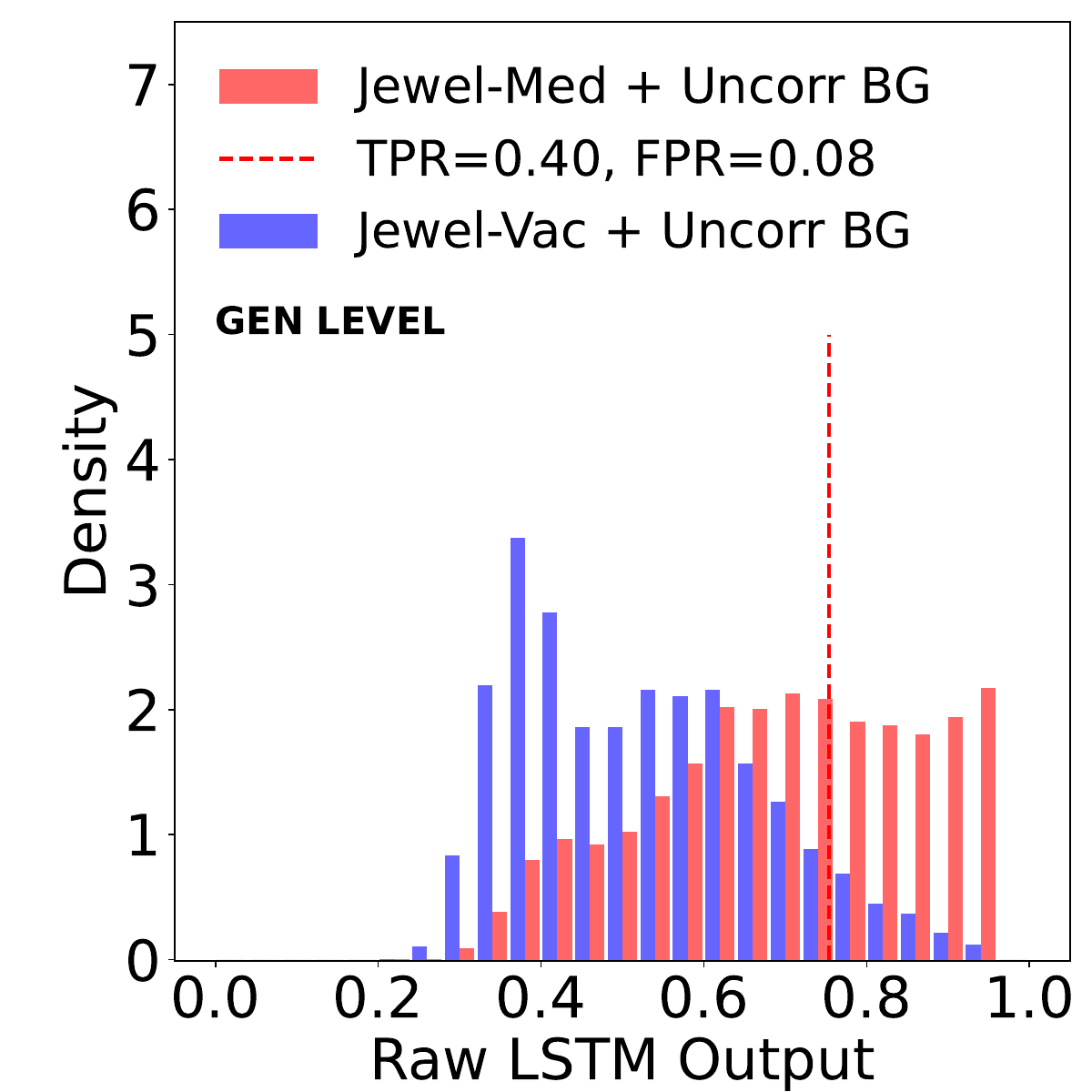}
            \caption{}
            \label{fig:LSTMgenYJ}
        \end{subfigure}
        \begin{subfigure}[b]{0.48\textwidth}
            \centering
            \includegraphics[width=\linewidth]{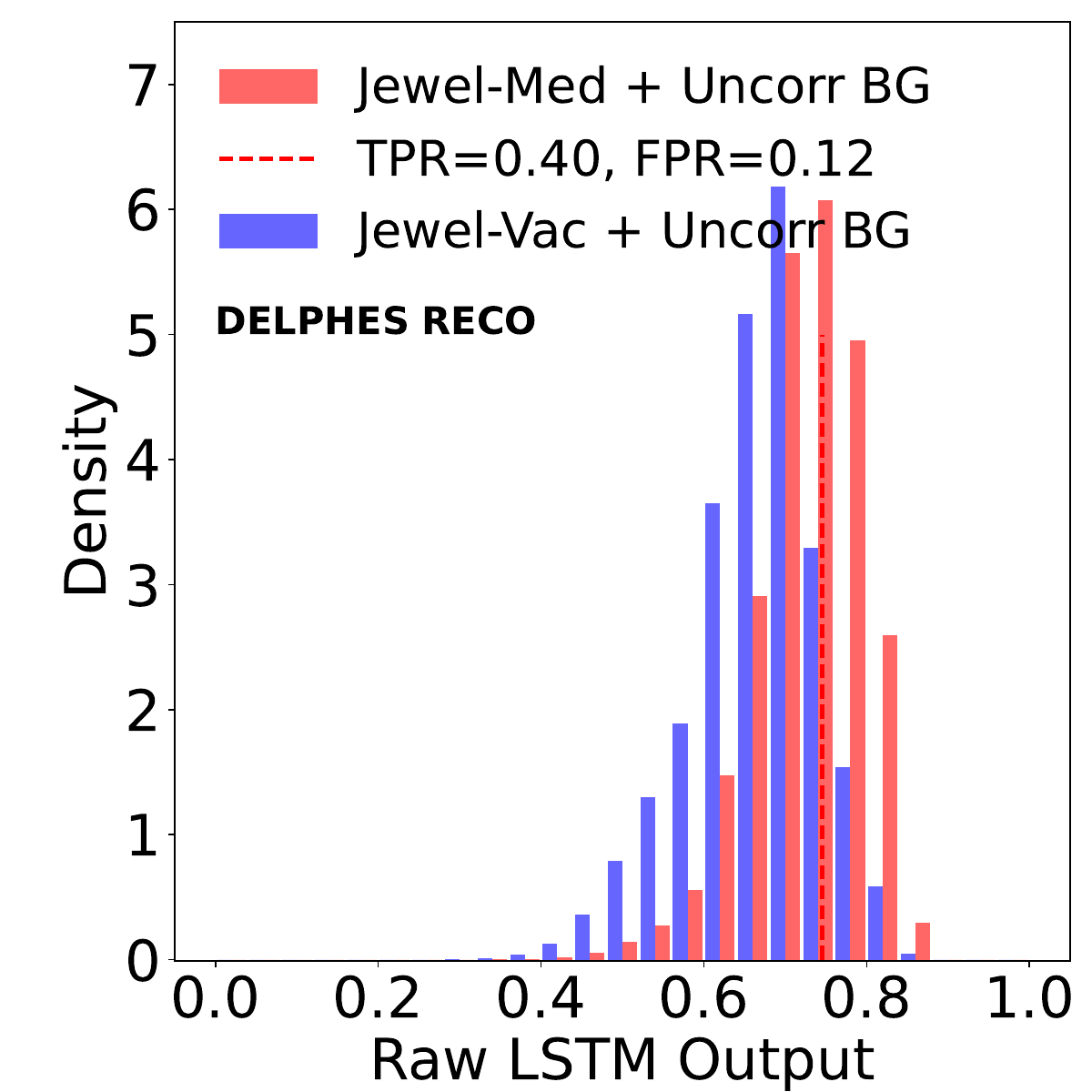}
            \caption{}
            \label{fig:LSTMeflowYJ}
        \end{subfigure}
        \centering
        \begin{subfigure}[b]{0.48\textwidth}
            \centering
            \includegraphics[width=\linewidth]{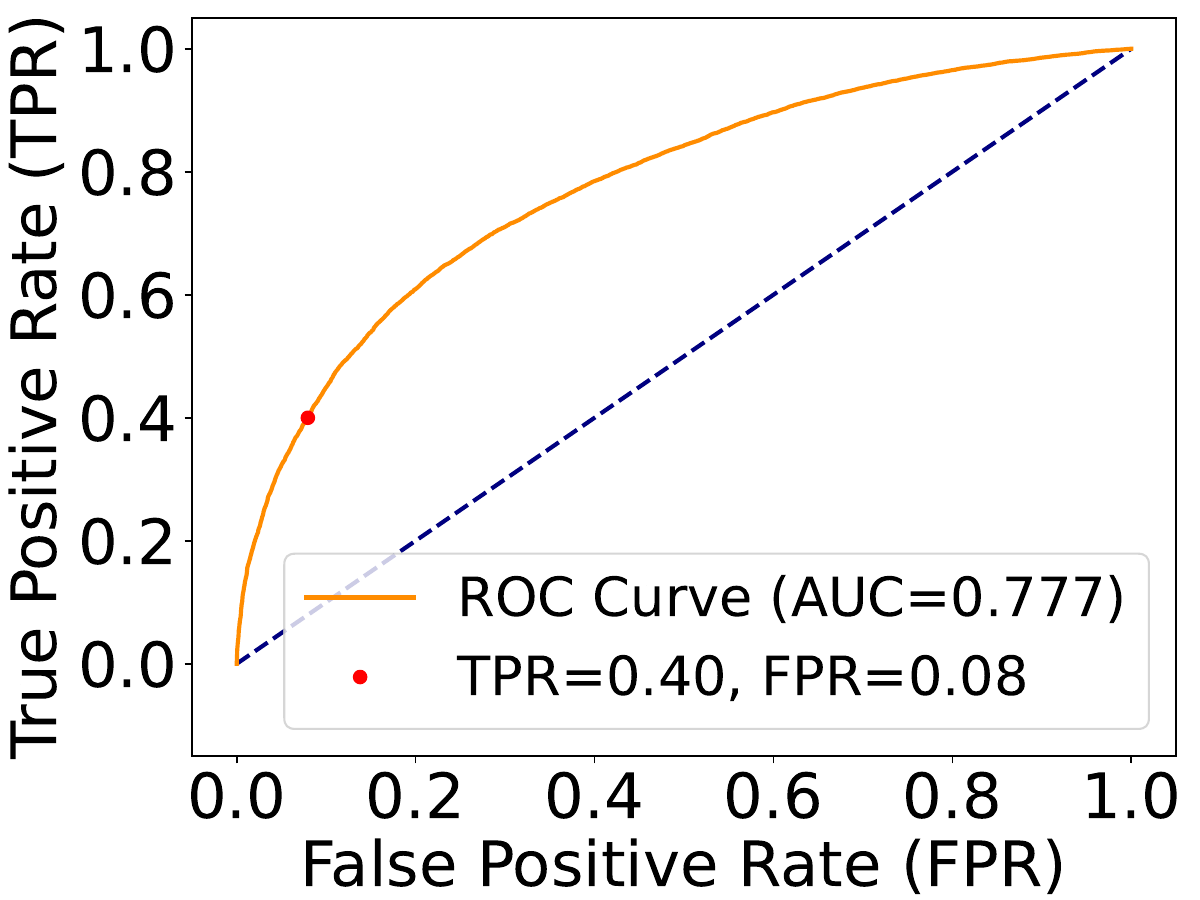}
            \caption{}
            \label{fig:ROCgenYJ}
        \end{subfigure}
        \begin{subfigure}[b]{0.48\textwidth}
            \centering
            \includegraphics[width=\linewidth]{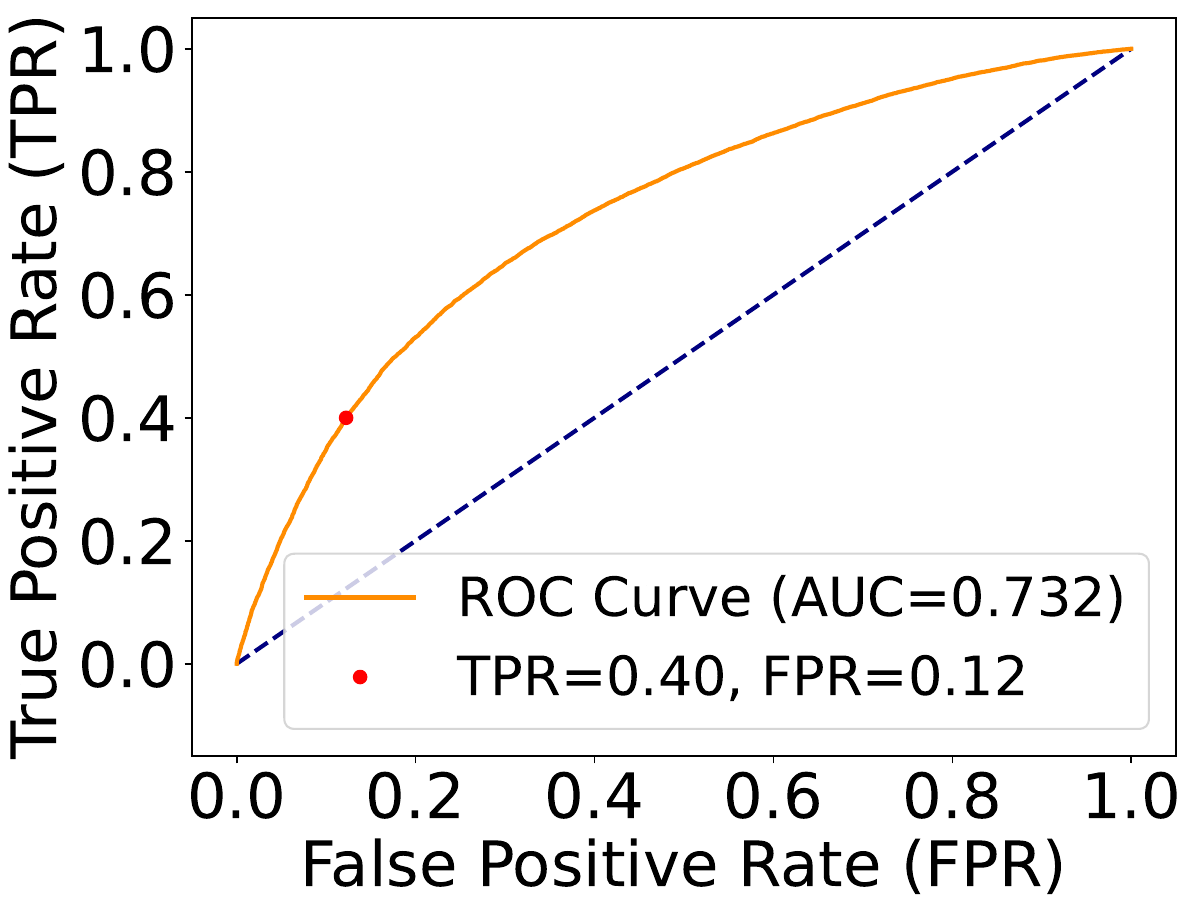}
            \caption{}
            \label{fig:ROCeflowYJ}
        \end{subfigure}
        \caption{Binary classification performance of the LSTM neural network at the GEN level (left) and RECO level (right). (a, b) Classification output; (c, d) ROC curve.}

        \label{fig:LSTMROC}
\end{figure}

The trained neural network can be used as a classifier to predict the probability that a jet is quenched. Figure~\ref{fig:LSTMROC} presents the distributions of the LSTM output and the receiver operating characteristic (ROC) curve. During training, the unquenched jets are labeled as 0, while the quenched jets are labeled as 1. The probability metric for jet quenching, learned during training, produces an output between 0 and 1 for each class.

\begin{figure}[h]
        \centering
        \begin{subfigure}[b]{0.33\textwidth}
                \centering
                \includegraphics[width=\linewidth]{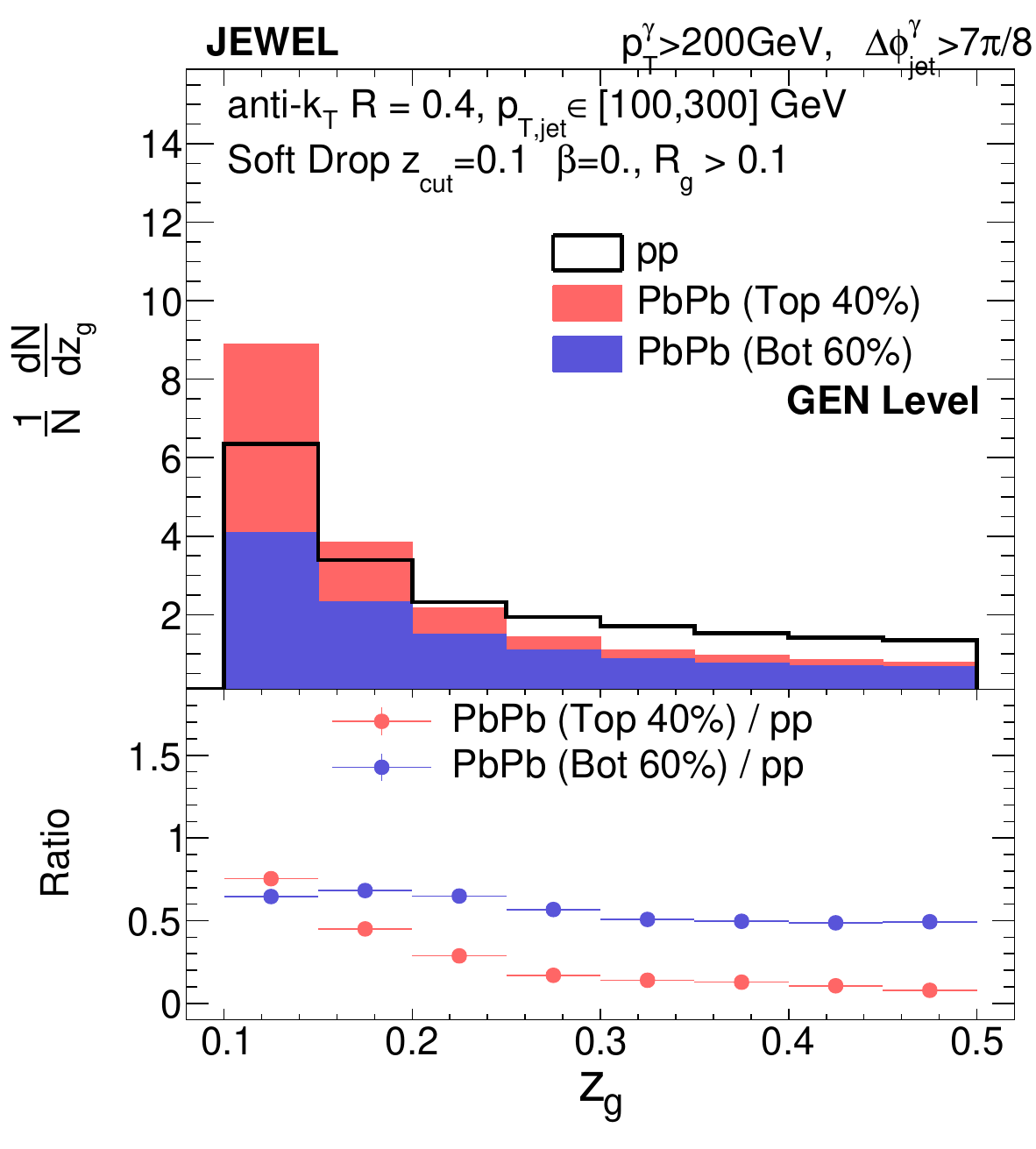}
                \caption{}
                \label{fig:NNJSzFeb}
        \end{subfigure}%
        \begin{subfigure}[b]{0.33\textwidth}
                \centering
                \includegraphics[width=\linewidth]{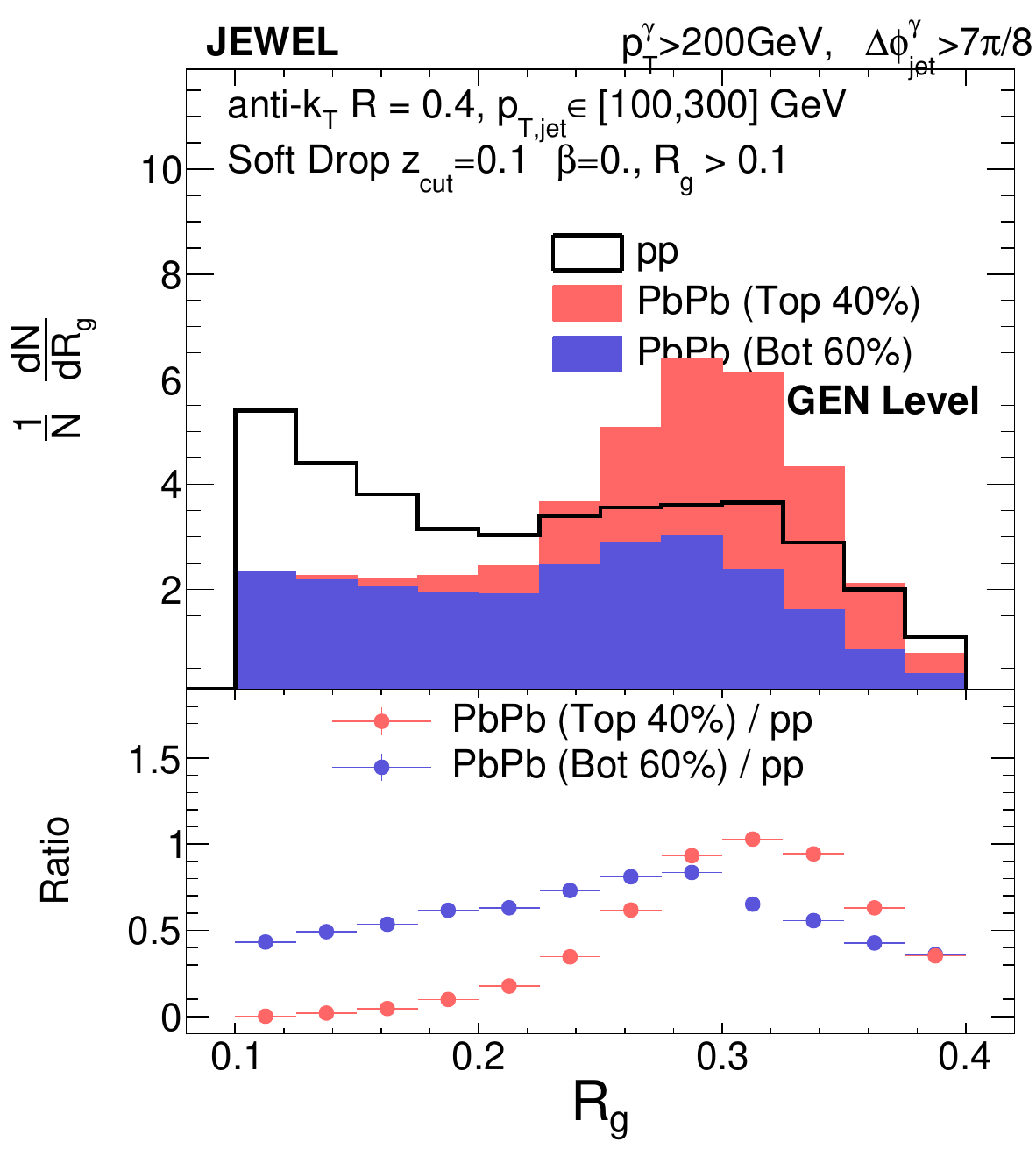}
                \caption{}
                \label{fig:NNJSdeltaGen}
        \end{subfigure}%
        \begin{subfigure}[b]{0.33\textwidth}
                \centering
                \includegraphics[width=\linewidth]{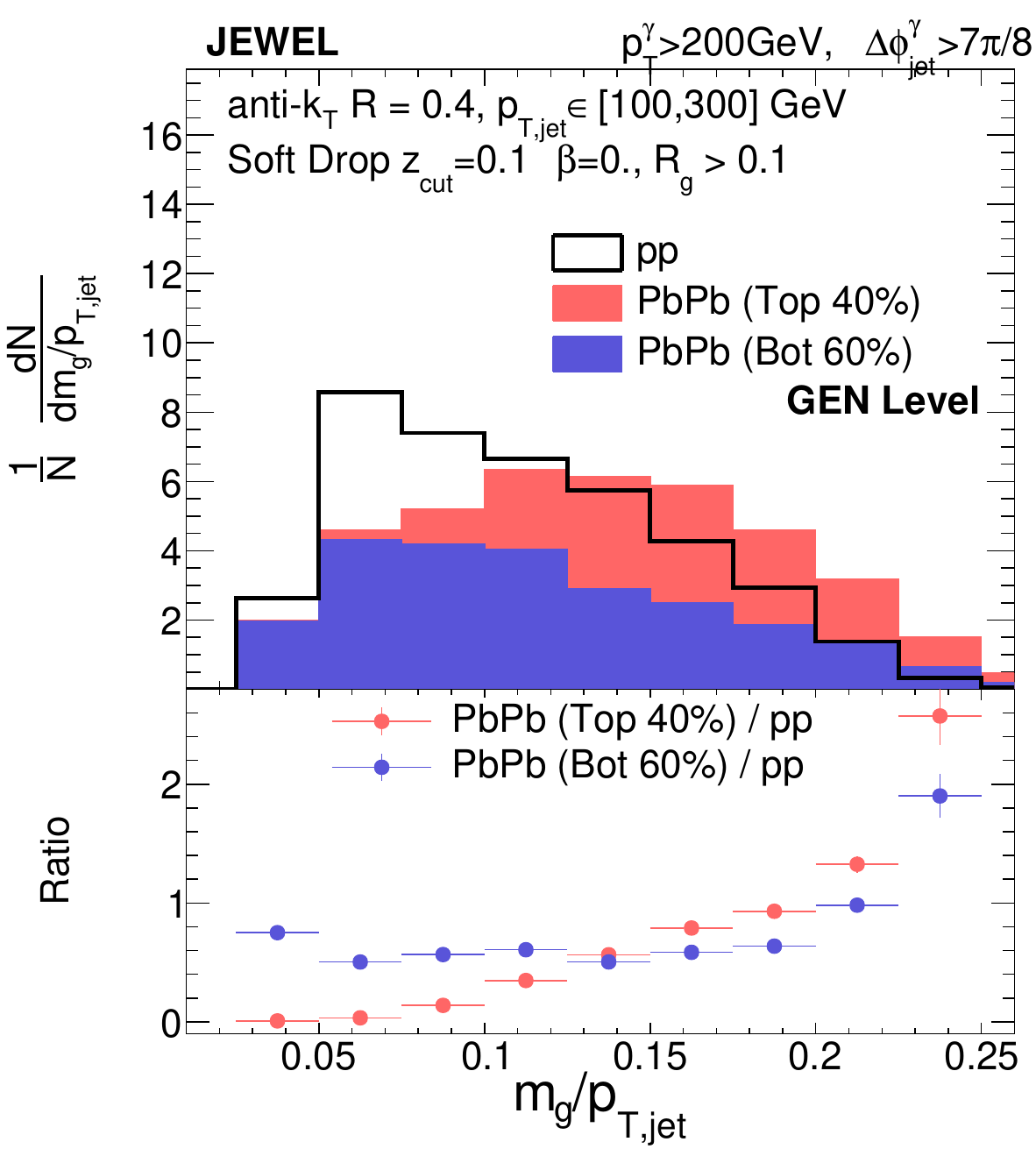}
                \caption{}
                \label{fig:NNJSmGen}
        \end{subfigure}%
        
        \centering
        \begin{subfigure}[b]{0.33\textwidth}
                \centering
                \includegraphics[width=\linewidth]{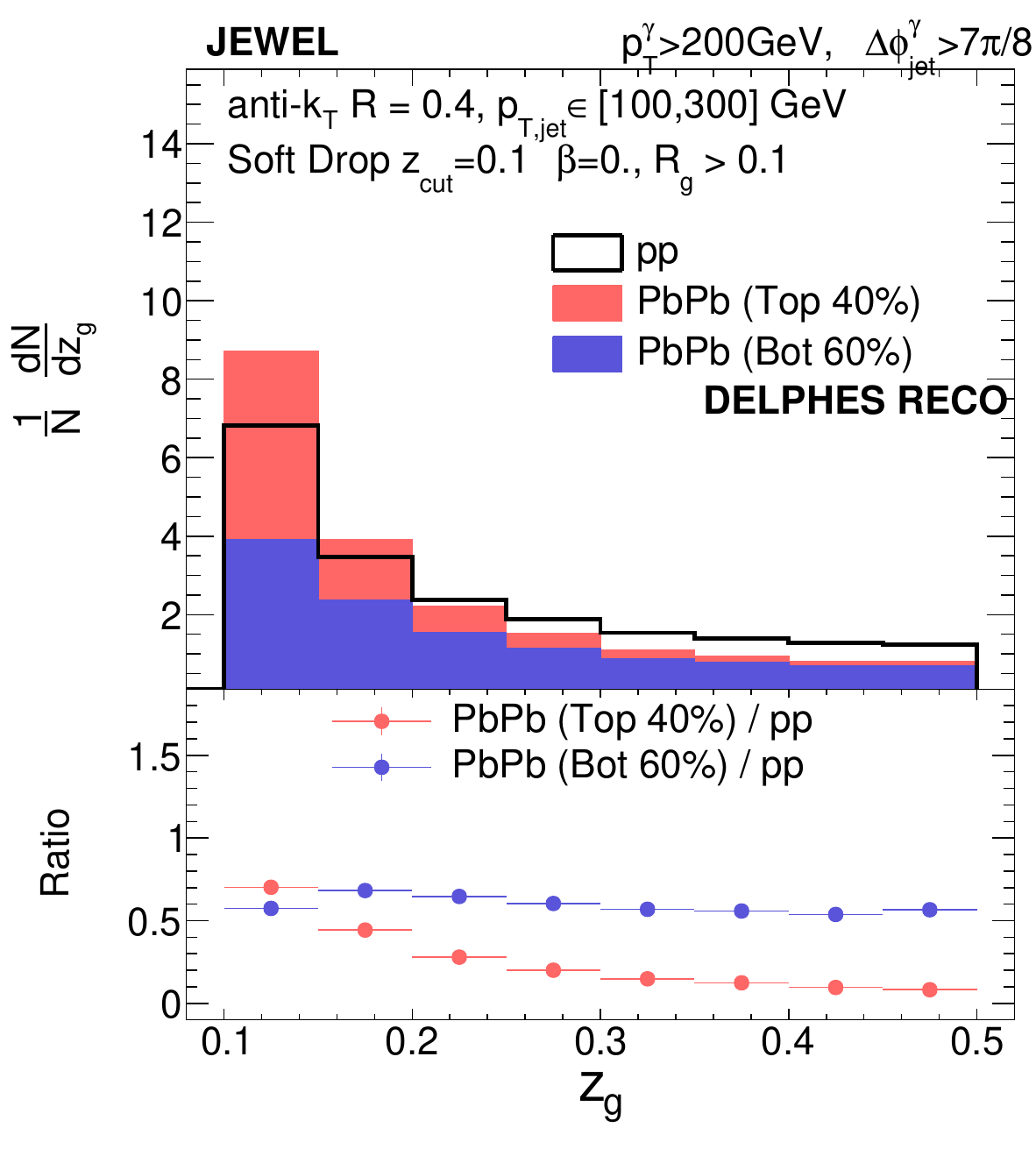}
                \caption{}
                \label{fig:NNzgeflow}
        \end{subfigure}%
        \begin{subfigure}[b]{0.33\textwidth}
                \centering
                \includegraphics[width=\linewidth]{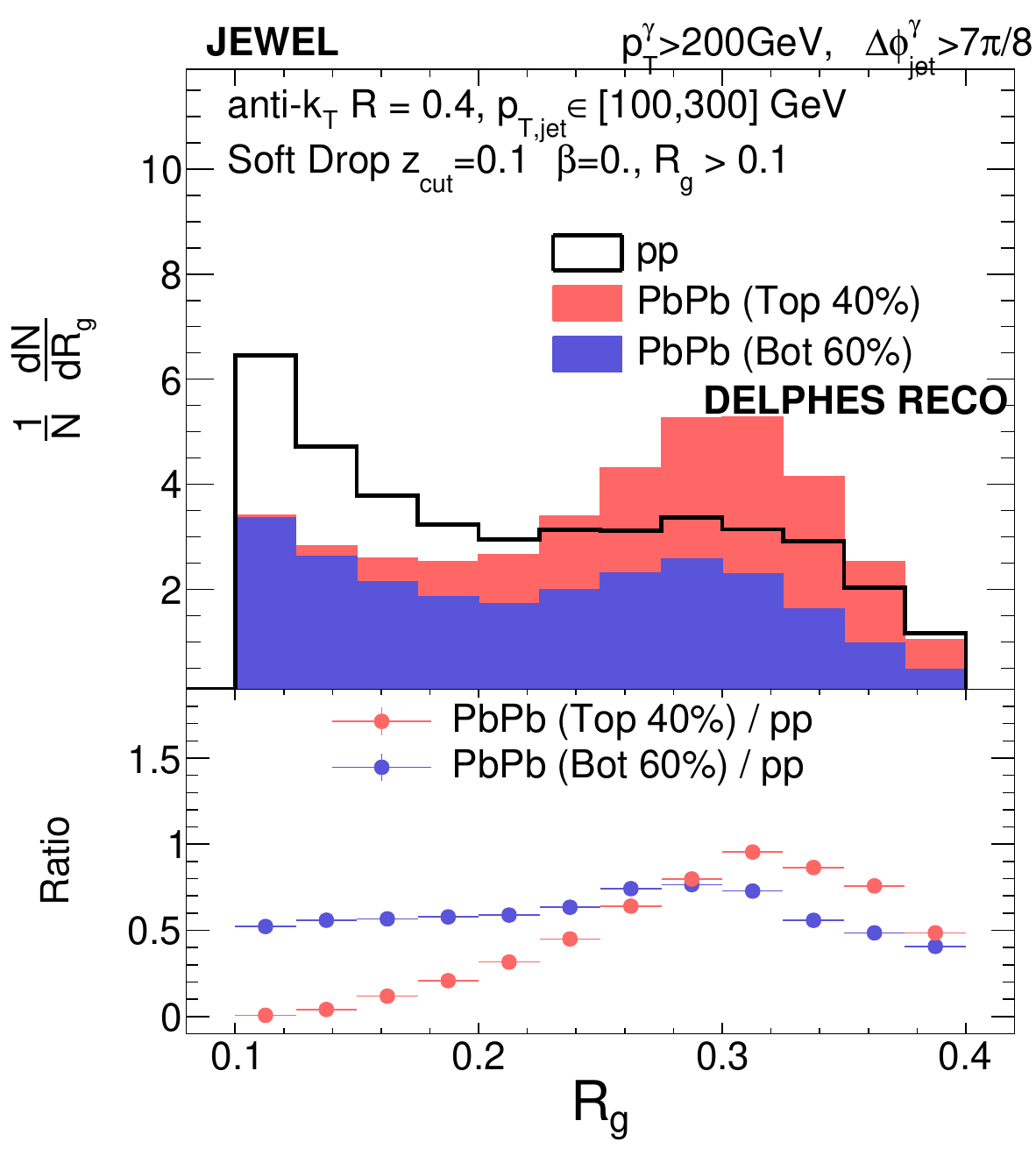}
                \caption{}
                \label{fig:NNdeltaeflow}
        \end{subfigure}%
        \begin{subfigure}[b]{0.33\textwidth}
                \centering
                \includegraphics[width=\linewidth]{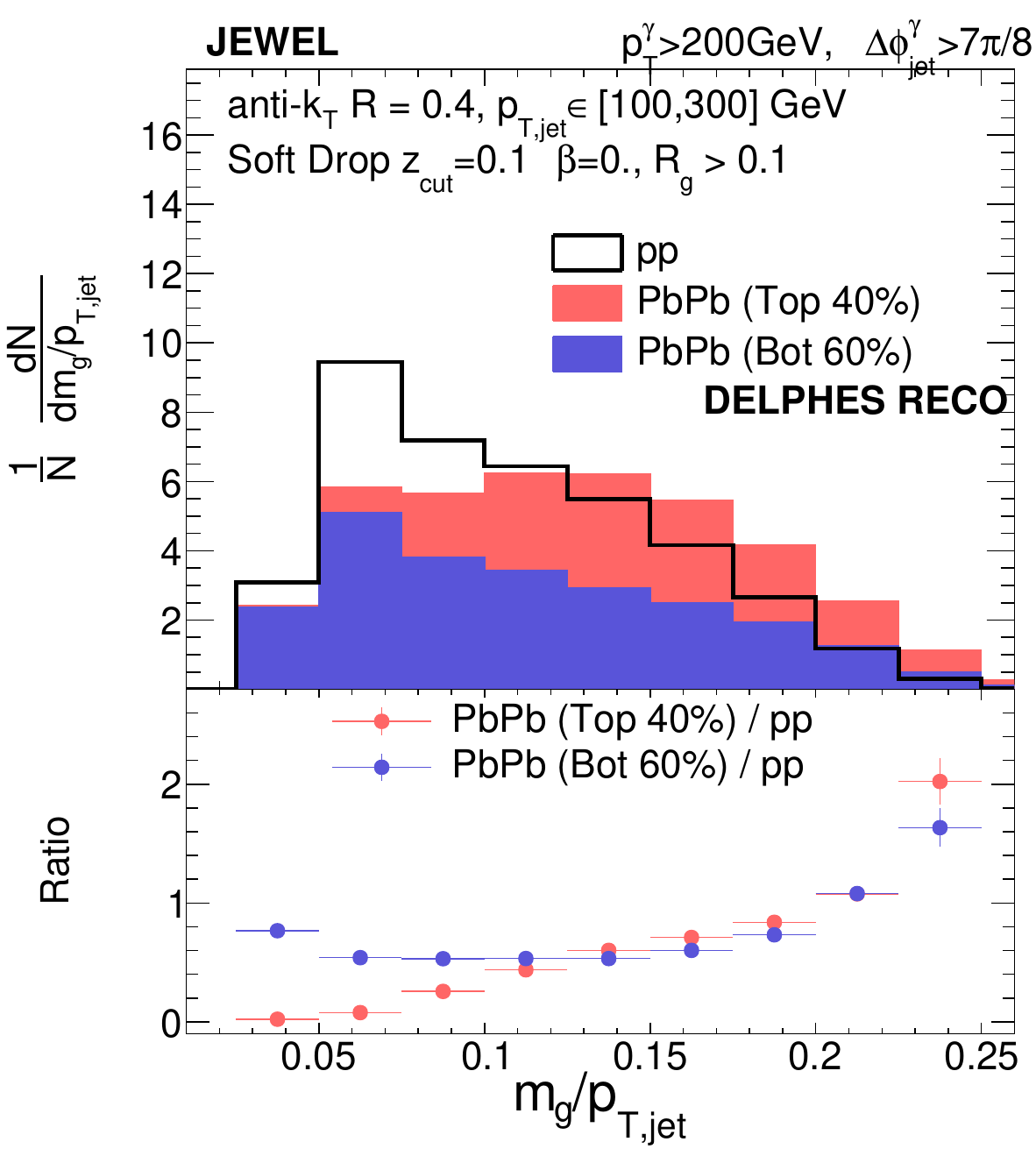}
                \caption{}
                \label{fig:NNmgeflow}
        \end{subfigure}%
        \caption{Neural network trained on GEN level (upper panels) and RECO level (lower panels) substructure variables of groomed jets, classifying quenched jets into two categories: top 40\% medium-like jets and bottom 60\% vacuum-like jets. (a, d) Momentum fraction (splitting function); (b, e) angular separation; (c, f) invariant jet mass.}

        \label{fig:NNmg}
\end{figure}

Samples from the quenched class (\textsc{Jewel}-Med) are divided into two subsets: the top 40\% and the bottom 60\%, based on the LSTM output. To explore the relationship between LSTM output and quenching effects, we present the distributions of groomed jet substructure variables in Fig.~\ref{fig:NNmg} for both subsets separately. For comparison, samples from the unquenched class (\textsc{Jewel}-Vac), serving as a baseline, are also included.  

The two subsets exhibit different levels of quenching effects. The top 40\% of the quenched class shows significant jet substructure modifications, with enhanced wider and softer splittings. In contrast, the bottom 60\% exhibits a quenching pattern more similar to the unquenched class. Neural networks trained at both the GEN and RECO levels yield consistent trends, indicating that despite detector smearing, the network effectively quantifies quenching effects. This shows that the machine learning approach successfully accounts for the detector effects on the inputs.

\clearpage
\subsection{Predictions for other observables}
\label{Predictions for other observables}
To better understand the relationship between the ML classifier results and the quenching effects, we apply the LSTM output to observables that were not included in the training. To study the quenching effects in detail, we assign a "quenching level" to each jet from 0 to 100\%, with 0\% being the most quenched, based on the LSTM output. The quenched sample is divided into five subsets equally. The most quenched class is denoted by Q 0-20\%, while the least quenched class is denoted by Q 80-100\%. 

\subsubsection{Photon-jet momentum imbalance}
\begin{figure}[h]
        \centering
        \begin{subfigure}[b]{0.48\textwidth}
            \centering
            \includegraphics[width=\linewidth]{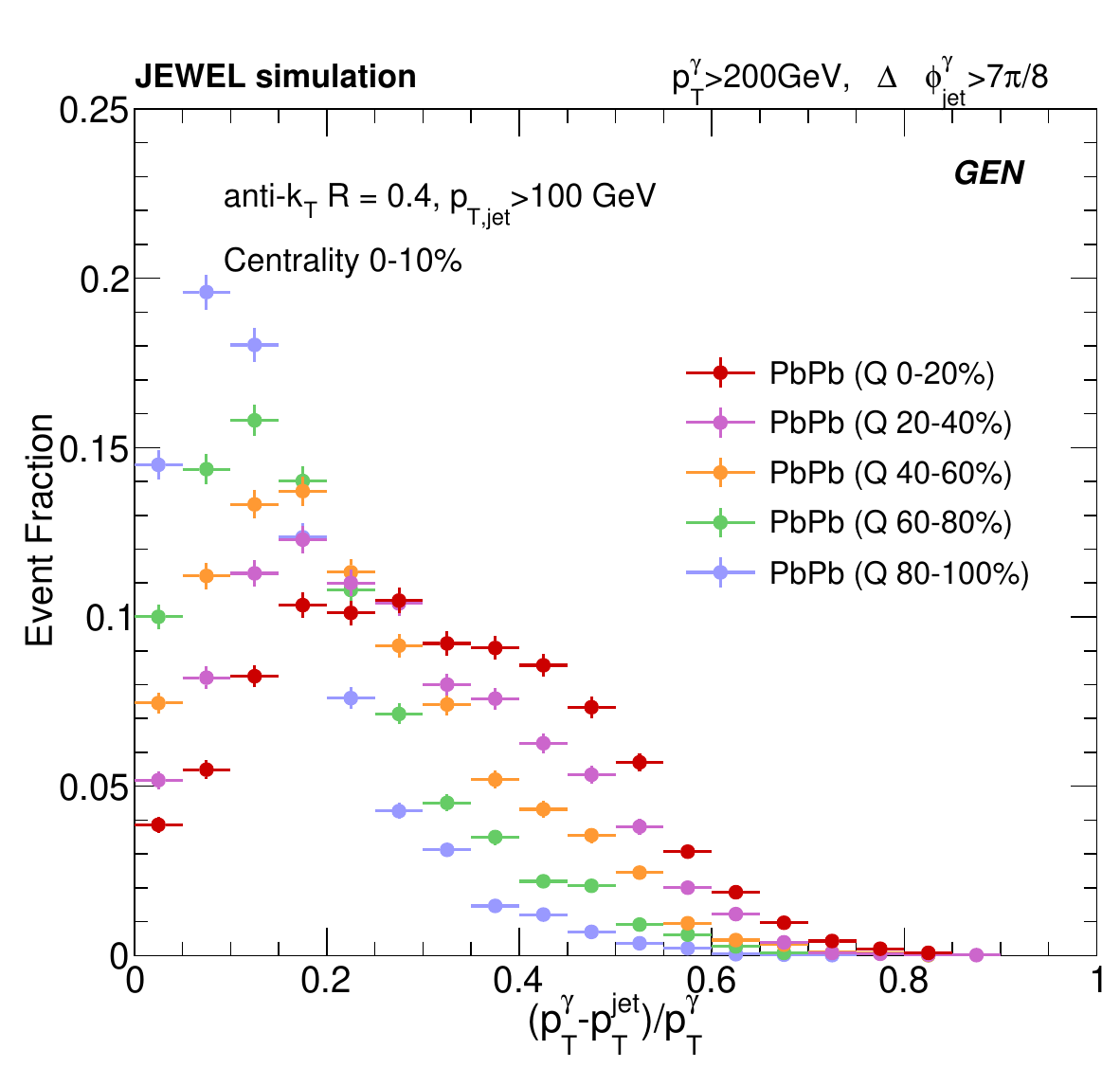}
            \caption{}
            \label{fig:PhotonJetImbGen_PbPb}
        \end{subfigure}
        \begin{subfigure}[b]{0.48\textwidth}
            \centering
            \includegraphics[width=\linewidth]{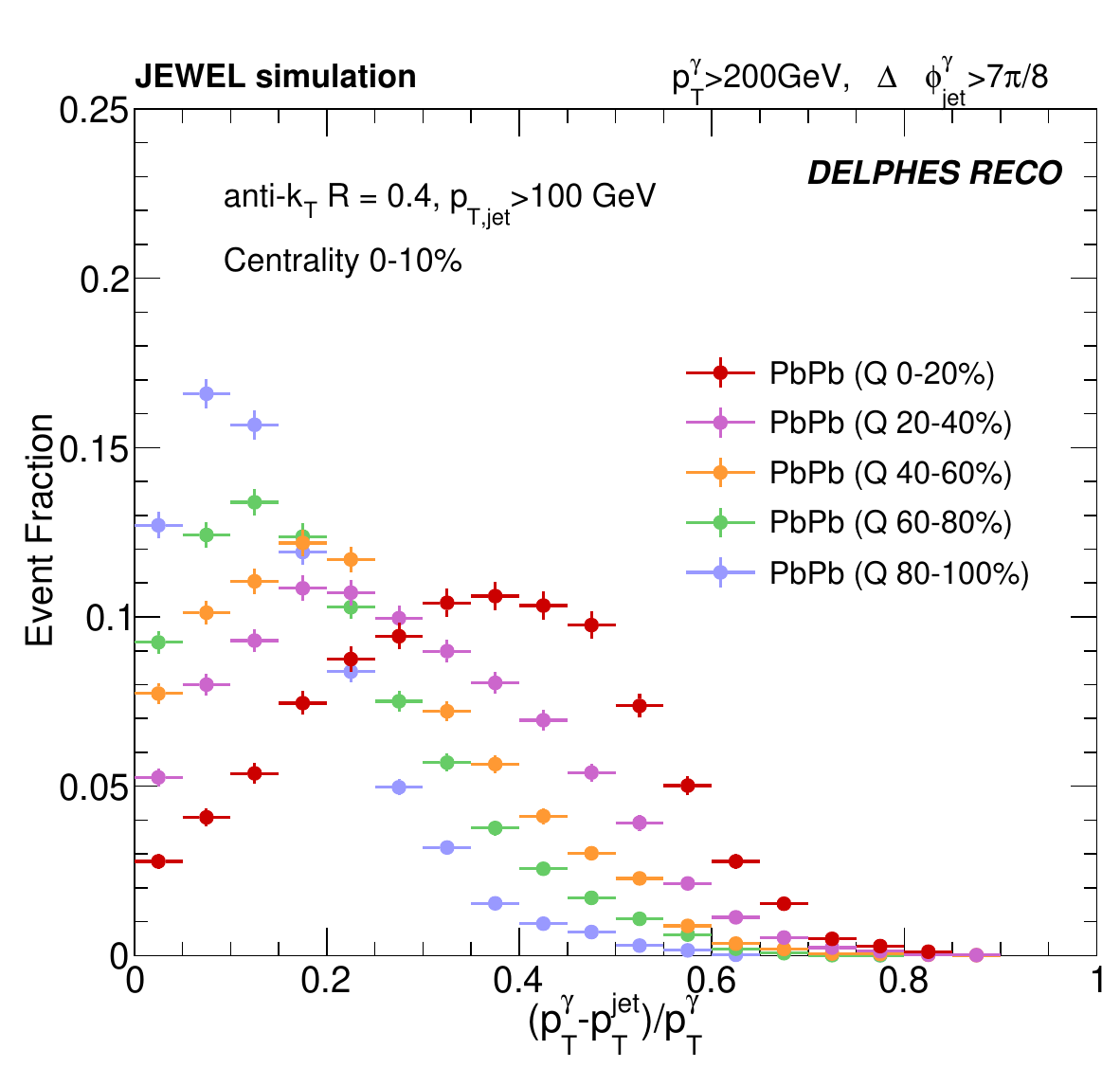 }
            \caption{}
            \label{fig:PhotonJetImbEflw_PbPb}
        \end{subfigure}
       \caption{Photon-jet momentum imbalance of five quenching classes from \textsc{Jewel}-Med (PbPb) identified by the LSTM classifier. (a) Five classes with different modifications compared to inclusive pp jets at the GEN level. (b) RECO level with detector effects simulated using \textsc{Delphes}.} 
        \label{fig:YJNN_PbPb}
\end{figure}

 For the five quenching classes, we measure the distributions of the photon-jet energy fraction at the GEN level in Fig.\ref{fig:PhotonJetImbGen_PbPb} and at the RECO level in Fig.\ref{fig:PhotonJetImbEflw_PbPb}. With or without detector effects, the most quenched class (Q 0-20\%) exhibits the greatest transverse momentum imbalance between the back-to-back jet and the photon. As the quenching level decreases toward the least quenched class (Q 80-100\%), the transverse momentum between the jet and the photon becomes more balanced, with the ratio distribution shifting closer to zero (no energy loss). The ordering in the peak positions of the five quenching classes remains consistent when the detector effects are included.

\subsubsection{Jet fragmentation function}

\begin{figure}[h]
        \centering
        \begin{subfigure}[b]{0.48\textwidth}
            \centering
            \includegraphics[width=0.9\linewidth]{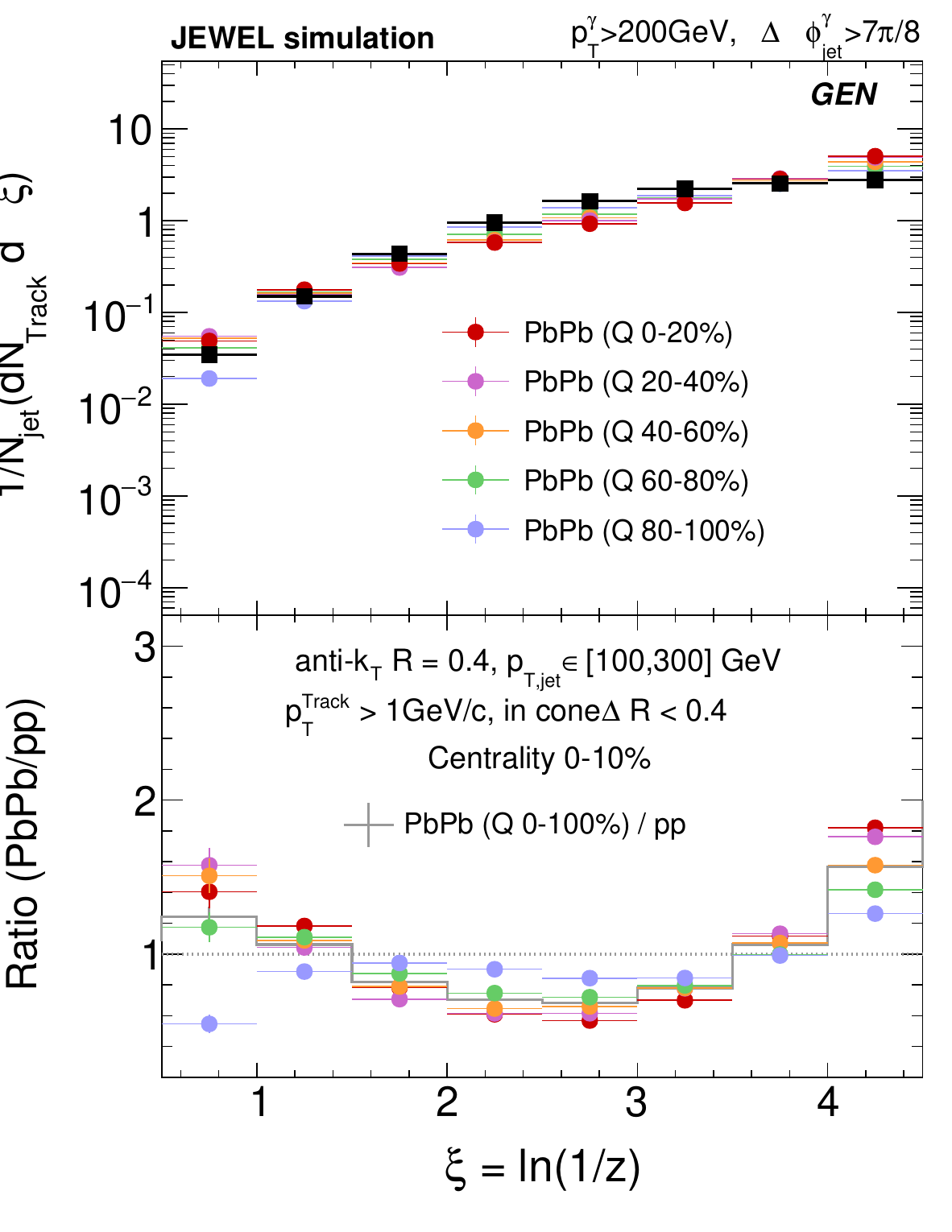}
            \caption{}
            \label{fig:JFFNNGen}
        \end{subfigure}
        \centering
        \begin{subfigure}[b]{0.48\textwidth}
            \centering
            \includegraphics[width=0.9\linewidth]{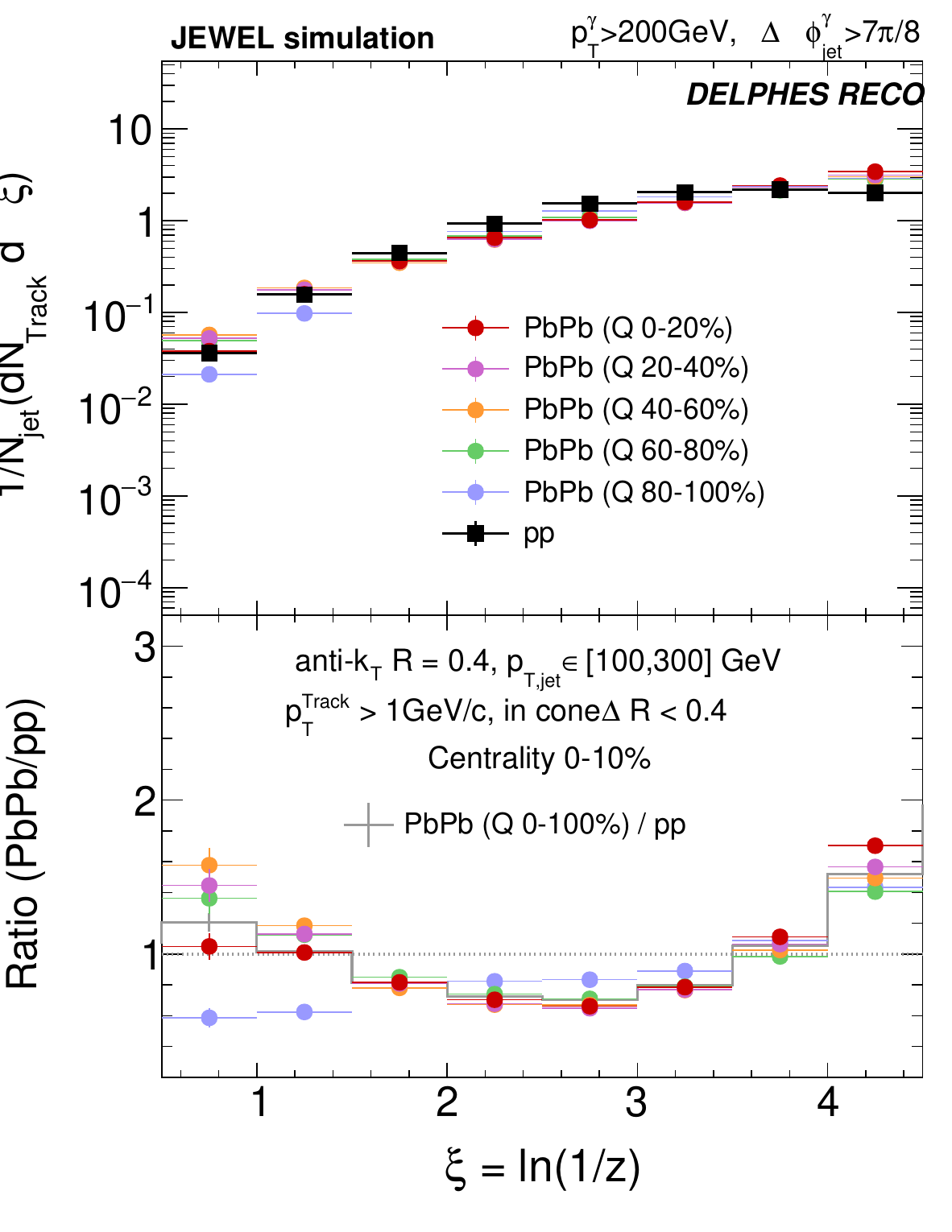}
            \caption{}
            \label{fig:JFFNNEFlw}
        \end{subfigure}
        
        \caption{Jet fragmentation function of five classes identified by the LSTM classifier. (a) Five classes of \textsc{Jewel}-Med (PbPb) jets, showing different modifications compared to inclusive pp jets at the GEN level. (b) RECO level with detector effects simulated using \textsc{Delphes}.}
        \label{fig:JFFNN_PbPb}
\end{figure}

The LSTM output is also applied to the jet fragmentation function,
\begin{equation}
    z=\frac{p^{\rm track}_{||}}{p^{\rm jet}}, \quad \xi=\ln\left(\frac{1}{z}\right),
\end{equation}
as shown in Fig.~\ref{fig:JFFNN_PbPb}. At both the GEN level (Fig.~\ref{fig:JFFNNGen}) and the RECO level (Fig.~\ref{fig:JFFNNEFlw}), jets with a stronger predicted quenching exhibit a stronger enhancement in the large $\xi$ region, corresponding to an increased contribution of soft particles within the jet cone. At the same time, strongly quenched jets (Q 0–20\%) show a greater depletion in the intermediate $\xi$ region compared to weakly quenched jets (Q 80–100\%), indicating increased energy loss.

Jet fragmentation functions exhibit class-dependent modifications relative to the baseline, represented by the ratio of inclusive \textsc{Jewel}-Med jets to inclusive \textsc{Jewel}-Vac jets. The class separation is clearly visible at GEN level and becomes less pronounced at RECO level. This behavior indicates that the underlying physical modification of the fragmentation pattern is preserved after reconstruction, with reduced separation due to detector effects and background fluctuations, particularly in the soft sector. 

\clearpage
\subsubsection{Jet momentum profile}

\begin{figure}[h]
        \centering
        \begin{subfigure}[b]{0.48\textwidth}
            \centering
            \includegraphics[width=0.9\linewidth]{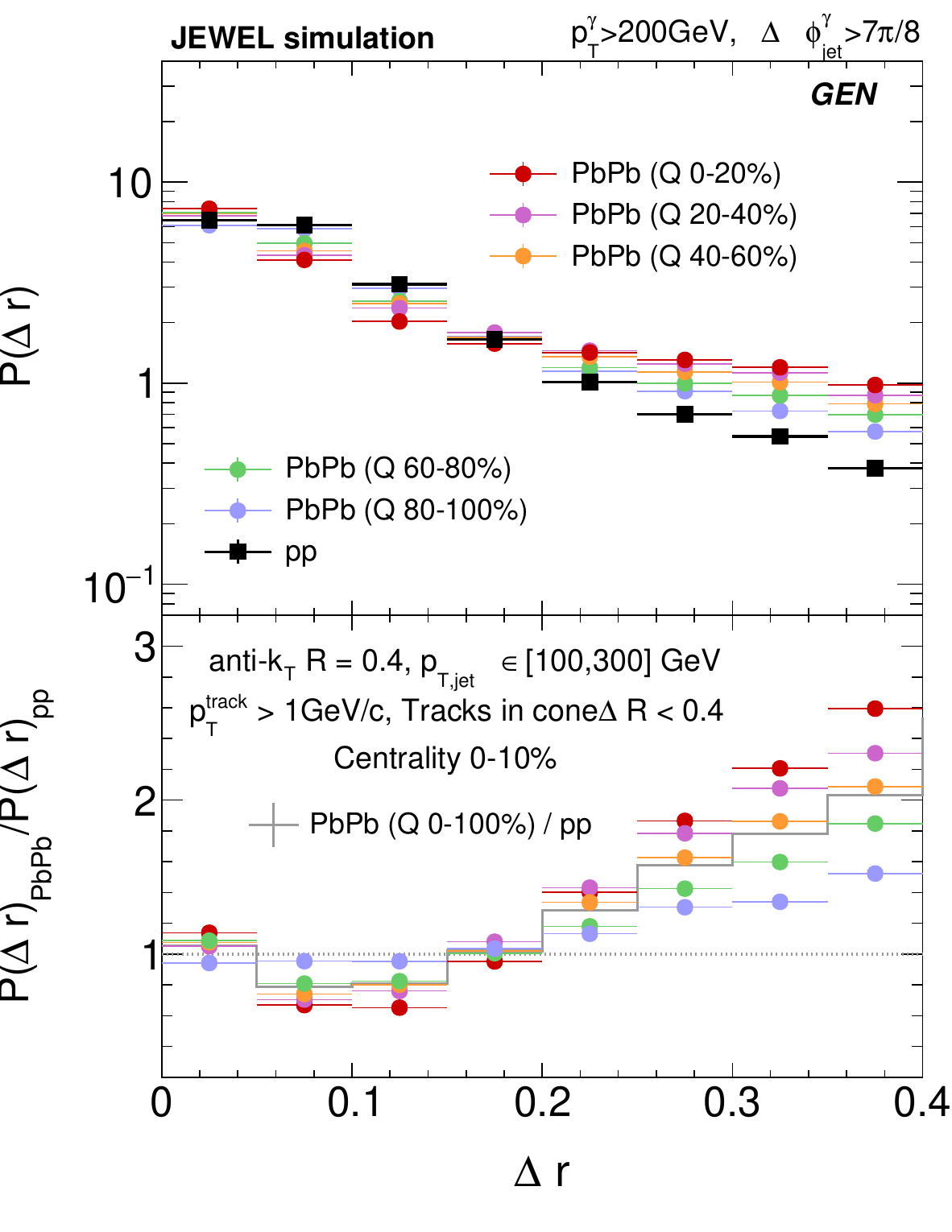}
            \caption{}
            \label{fig:JSNNGen}
        \end{subfigure}
        \begin{subfigure}[b]{0.48\textwidth}
            \centering
            \includegraphics[width=0.9\linewidth]{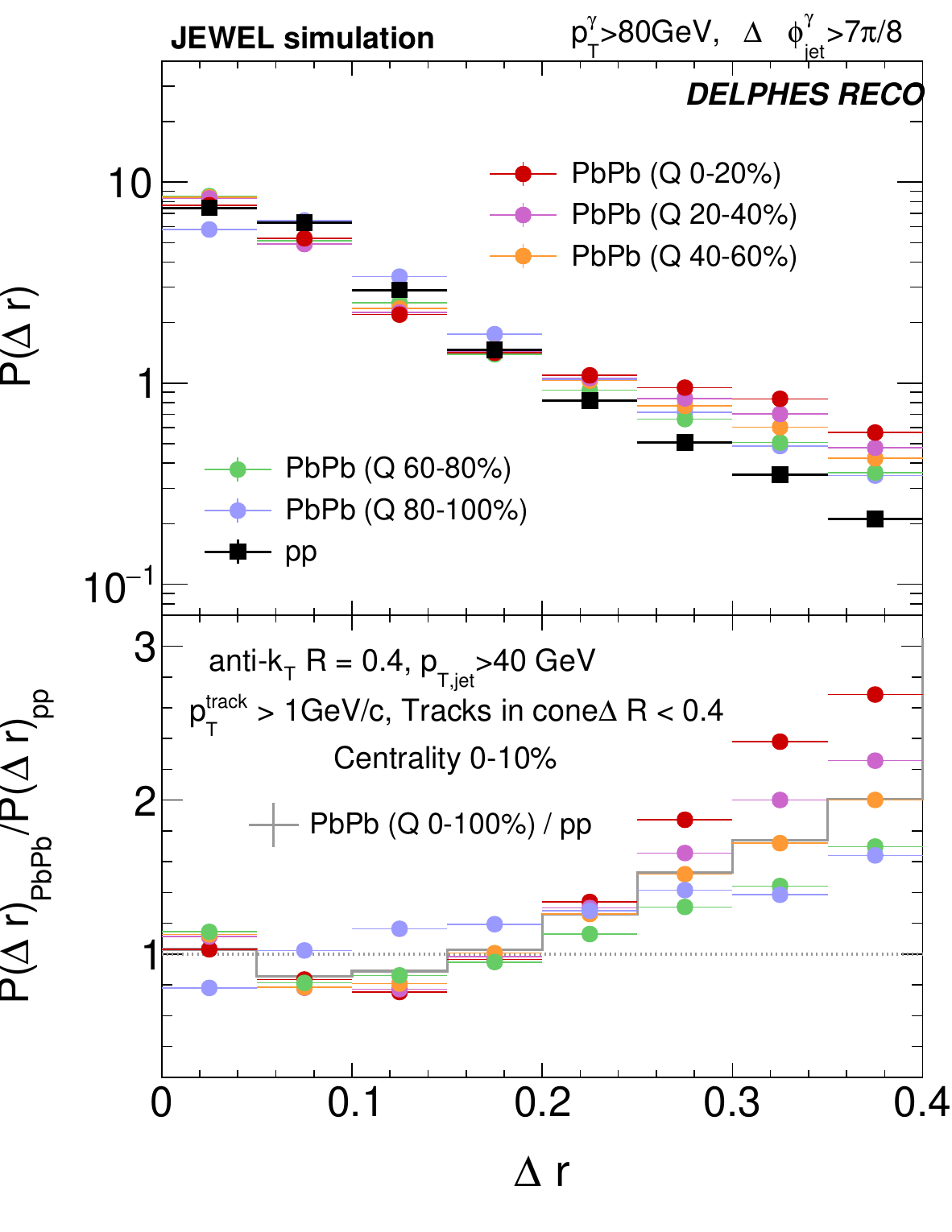 }
            \caption{}
            \label{fig:JSNNEFlw}
        \end{subfigure}
        \caption{Jet momentum profile of five classes with different quenching levels identified by the LSTM neural network. (a) Five quenching classes for the \textsc{Jewel}-Med (PbPb) jet sample at the GEN level. (b) RECO level with detector effects simulated using \textsc{Delphes}.}

        \label{fig:JSNN}
\end{figure}

ML classification results for the jet momentum profile (integrated jet shape),
\begin{equation}
P(r)=\frac{1}{\delta r}\frac{1}{N_{\rm jet}}\sum_{\rm jets}{\sum_{{\rm tracks}\in[r_a,r_b)}p^{\rm track}_T},
\end{equation}
are shown in Fig.~\ref{fig:JSNN}. At both the GEN level (Fig.~\ref{fig:JSNNGen}) and the RECO level (Fig.~\ref{fig:JSNNEFlw}), jets with a stronger predicted quenching exhibit enhanced energy at larger $\Delta R$ and the corresponding depletion at smaller $\Delta R$. This pattern is consistent with a redistribution of jet energy toward softer particles at larger angles from the jet axis in strongly quenched jets. 

In contrast, the least quenched jets (Q 80–100\%) behave similarly to vacuum jets, with their ratios to inclusive pp jets showing a relatively flat dependence on $\Delta R$ compared to more strongly quenched classes.

\clearpage

\section{Conclusion}
\label{sec:Conclusion}

We demonstrated that a well-trained ML classifier based on sequential jet substructures can effectively distinguish true quenching features from other sources that may mimic quenching, such as thermal background and detector effects. To validate this, we cross-checked observables not included in the training process, e.g. photon-jet momentum imbalance, jet fragmentation function, and jet momentum profile. We found that the jets identified by the ML classifier with varying quenching levels exhibit the corresponding modifications in these observables.

Using the \textsc{Delphes} simulation of the CMS detector response and particle flow candidate equivalents, we predicted the results of applying the ML method to data analysis. The results indicated that, even with detector effects, our ML classifier can still effectively learn from quenching features, consistent with the jet energy loss estimated from the photon energy. We demonstrated significant potential of the machine learning approach for application to experimental data, where it can analyze a wide range of jet observables on a jet-by-jet basis. Such studies will help disentangle a variety of competing mechanisms that lead to jet modifications in QGP.


\acknowledgments
The authors thank R. Kunnawalkam Elayavalli and J. Viinikainen for insightful discussions. The authors also acknowledge the support of the Vanderbilt ACCRE computing facility. This work was supported by US Department of Energy Grant No. DE-FG05-92ER40712. 

\clearpage

\bibliographystyle{JHEP}
\typeout{}
\bibliography{JetML}
\end{document}